\pdfoutput=1
\documentclass[iop,twocolumn,jphysc,8pt,showpacs,floatfix,nofootinbib,
superscriptaddress
]{revtex4-1}
\usepackage{subfigure}
\usepackage{amssymb}
\usepackage{amsfonts}
\usepackage{amsmath}
\usepackage{amsthm}
\usepackage{epsfig}
\usepackage{graphicx}
\usepackage[usenames,dvipsnames]{color}
\usepackage[latin1]{inputenc}
\usepackage{hyperref}
\usepackage{comment}
\begin{document}

\title{Origin of the literature discrepancies in the fractional reduction of the apex-field enhancement factor considering small clusters of field emitters}

\author{Thiago A. de Assis}
\email{thiagoaa@ufba.br}
\address{Instituto de F\'{\i}sica, Universidade Federal da Bahia,
   Campus Universit\'{a}rio da Federa\c c\~ao,
   Rua Bar\~{a}o de Jeremoabo s/n,
40170-115, Salvador, BA, Brazil}

\author{Fernando F. Dall'Agnol}
\address{Department of Exact Sciences and Education (CEE), Universidade Federal de Santa Catarina,
 Campus Blumenau,
   Rua Pomerode 710 Salto do Norte,
Blumenau 89065-300, SC, Brazil}
\email{fernando.dallagnol@ufsc.br}

\begin{abstract}

Numerical simulations are important when assessing the many characteristics of field emission related phenomena. In small simulation domains, the electrostatic effect from the boundaries is known to influence the calculated apex field enhancement factor (FEF) of the emitter, but no established dependence has been reported at present. In this work, we report the dependence of the lateral size, $L$, and the height, $H$, of the simulation domain on the apex-FEF of a single conducting ellipsoidal emitter. Firstly, we analyze the error, $\varepsilon$, in the calculation of the apex-FEF as a function of $H$ and $L$. Importantly, our results show that the effects of $H$ and $L$ on $\varepsilon$ are scale invariant, allowing one to predict $\varepsilon$ for ratios $L/h$ and $H/h$, where $h$ is the height of the emitter. Next, we analyze the fractional change of the apex-FEF, $\delta$, from a single emitter, $\gamma_1$, and a pair, $\gamma_2$. We show that small relative errors in  $\gamma_1$ (i.e., $\varepsilon\approx0.5\%$), due to the finite domain size, are sufficient to alter the functional dependence $\delta(c)$, where $c$ is the distance from the emitters in the pair. We show that $\delta(c)$ obeys a recently proposed power law decay [R. G. Forbes, J. of Appl. Phys. \textbf{120}, 054302 (2016)], at sufficient large distances in the limit of infinite domain size ($\varepsilon=0$, say), in contrast to a long time established exponential decay [J.-M. Bonard \textit{et al.} Advanced Materials \textbf{13}, 184 (2001)]. We also shown that this functional dependence is respected for various systems which includes infinity arrays and small clusters of emitters with different shapes. Thus, power law functional dependence, $-\delta \sim c^{-m}$, with $m=3$, is suggested to be a universal signature of the charge-blunting (CB) effect in small clusters or arrays, at sufficient large distances between emitters with any shape. These results explain the origin of the discrepancies in the literature and improves the scientific understanding of the field electron emission theory, for accurate characterization of emitters in small clusters or arrays. Finally, our results reinforce that the consequences of CB for a small cluster of emitters are also expected for infinity arrays.

\end{abstract}

\maketitle

\section{Introduction}

Cold field electron emission (CFE) by a conducting surface, when a strong
electrostatic field is applied, is a phenomenon that has led to scientific and technological developments \cite{Muller1937,Muller1951,Muller1956,Edgcombe,Cole2015chapter,PRE1,PRLNanotube,PRL2,PRB1,Jeffreys,FowlerN,Burgess,MG,Forbes,ForbesJPhysA,Forbes2013}. Recently, great attention has been given by field emission community to the research of the electrostatics behind single or small cluster of emitters \cite{Jensen2015,JensenAPL2015,Harris2015AIP,Harris2016,Jensen2016AIPA,Harris2016,RFJAP2016,deAssis1,deAssisJAP,ForbesAssis2017,FT2017JPCM,Jensen2017}, aiming to understand the emitter's interaction that leads to charge-transfer and neighbor-field effects \cite{RFJAP2016}. Particularly, an important Field Emission (FE) characterization parameter is the apex-field enhancement factor (FEF) \cite{Edgcombe}. The apex-FEF, for a single emitter in a parallel-plate diode configuration, $\gamma_{1}$, can be defined by,
\begin{equation}
\gamma_{1} = E_{1}/E_{M},
\label{FEFa}
\end{equation}
where $E_{1}$ corresponds to the local field at the apex of the single emitter and $E_{M}$ the applied (macroscopic) field. In the case of a pair of identical emitters, the corresponding apex-FEF is expected to be reduced to a lower value, $\gamma_{2}$, mainly due to the physical effect called charge-blunting (also known by shielding effect), which results from the requirement that the Fermi level be the same everywhere in the emitter plate and both emitters. Thus, it is possible to define a fractional ratio $\rho \equiv \gamma_{2}/\gamma_{1}$ and a fractional change in the apex-FEF, $\delta$, as

\begin{equation}
\delta \equiv \rho - 1 = \frac{\gamma_{2} - \gamma_{1}}{\gamma_1}.
\label{delta21}
\end{equation}
With infinite regular arrays, this fractional difference has been extensively analyzed. In a recent work \cite{RFJAP2016}, Forbes used floating spheres at emitter-plate potential (FSEPP) model to pointed out that discrepancies exist between the analytical results and formula previously fitted to numerical results in the literature \cite{Bonard2001,Jo}, based on finite elements (for one single emitter apex-FEF calculations - using the computer program CIELAS2 - Granta Electronics, Cambridge, U.K. \cite{Jo}) or finite differences (for calculations related to $\delta$ in an array - by using the program MACSIMION (version 2.04) \cite{Jo}) techniques, which requires a finite domain for simulation. These discrepancies include the functional dependence between $\delta$ and the separation between emitters. It is worth mentioning that the performance of the field emitters is affected by space charge due to
the sensitive dependence of the emitted current on the local FEF \cite{Refnew2}. Predictions of the impact of space charge on the emitted current has been made by using line charge distribution \cite{LCMnew}. However, space-charge effects only come into significant play at relatively high fields and current densities. In this work the treatment, like almost all other treatments in the literature, applies at fields below the fields at which space-charge becomes important.

For two conducting floating spheres, Forbes showed exactly that the fractional apex-FEF decays according as $-\delta \sim c^{-3}$, for large $c$, where $c$ corresponds to the distance between the centers of the spheres \cite{RFJAP2016}. These results had been confirmed recently using finite elements method for a pair of conducting identical hemispheres on cylindrical post (HCP) \cite{ForbesAssis2017}, where $c$ in this case, and as used in this work, corresponds to the distance between the symmetry axis of the emitters. This suggest that the power law functional dependence $-\delta \sim c^{-3}$ may be more general and applicable to other emitters with different shapes. This present work investigates this possibility. However, the well known Bonard \textit{et al.} fitting formula \cite{Bonard2001}, as reformulated by Jo \textit{et al.} \cite{Jo}, shows that $\delta$ in a small array of HCP conducting emitters \cite{Bonard2001,Jo} is given by

\begin{equation}
-\delta^{n} = \exp\left\{-2.3172 \left(c/h\right)\right\},
\label{delta21Bonard}
\end{equation}
where $h$ is the height of the emitters. Several experimental works have adopted Eq.(\ref{delta21Bonard}) or previous dimensionally inconsistent equation shown in the Ref.\cite{Bonard2001} to understand the effects produced by shielding in an array of emitters. See, for example, Refs. \cite{SRZhang,ZhuRef,Maiti,Sheini,WangAPL}. The superscript ``$n$" from hereon, shall indicate that the quantity carrying this index was determined numerically, opposed to the same variables without the index, which is determined exactly.

Generalization of the Eq.(\ref{delta21Bonard}) has been recently proposed by Harris and collaborators by using line charge model (LCM) \cite{Harris2015AIP,Harris2016}. They obtained good agreement with LCM-generated data [i.e., curves of $\gamma(c)$ ($\beta(b)$ in their original notation)] using a two-parameters fitting formula, similar to Eq.(\ref{delta21Bonard}), written as

\begin{equation}
-\delta^{LCM} = \exp\left\{a \left(c/h\right)^{\kappa}\right\},
\label{delta21Jensen}
\end{equation}
where $a$ is a fitting parameter different, in general, from $-2.3172$ and $\kappa$ is an additional ad-hoc parameter that improves the fitting in a specific region of their interest. At this point, it is important to point out that our main purpose in this work is not compare our results with those from LCM calculations, where there is no finite domain as required by simulations using finite elements or finite differences methods.
Our main goal is to investigate the origin of the discrepancies observed in literature, which propose an exponential-type fitting, originally suggested by Bonard \textit{et al.} and Jo \textit{et al.} \cite{Bonard2001,Jo}, based on a previous works from Nilson \textit{et al.} \cite{Groning1} and Gr\"{o}ning \textit{et al.} \cite{Groning2000} numerical Laplace simulations.

Furthermore, in Refs. \cite{Jo,Groning2000}, the authors employed numerical calculations based on finite differences method to evaluate the electric field distribution using the program MACSIMION (version 2.04) \cite{DAHL20003}. In these cases, the functional dependence of $\delta(c)$ for large $c$ may be appear if the size of the simulation domain is not large enough, as will be shown. The sizes of the simulation domains were not mentioned in Refs. \cite{Jo,Groning2000}. Probably, the starting point of the authors of Refs. \cite{Harris2015AIP,Harris2016} for using the exponential fitting given by Eq.(\ref{delta21Jensen}) was the results obtained from original works \cite{Bonard2001,Jo}, since an analytical formula for $-\delta^{LCM}$ was not reported in the present stage of knowledge. Also, we verified that the power law decay observed in the analytical solution of the FSEPP model is also present in several other geometries, under sufficient numerical precision. We show that $-\delta^{n} \rightarrow c^{-3}$ as $c\rightarrow \infty$ for several geometries. As we will show, in this limit one parameter power law decay fitting formula is confirmed, in contrast with exponential fitting functions.

This paper is organized as follows. In Section II, the results from numerical simulations of a single-emitter apex-FEF, $\gamma_{1}^{n}$, are presented. The results of the fractional
change in the apex-FEF from two identical emitters case, $\delta^{n}$, are presented in Section III. In Section IV, we discuss the universality of the functional relation $-\delta \sim c^{-m}$, with $m=3$, as $c\rightarrow \infty$, corroborated by precise numerical calculations applied to various electrostatic systems, from small clusters to infinite arrays. Finally, Section V present main conclusions of this work.

\section{Apex-FEF for a single emitter case}
\label{SecII}

\begin{figure}[h!]
\includegraphics [width=6.6cm,height=5.6cm] {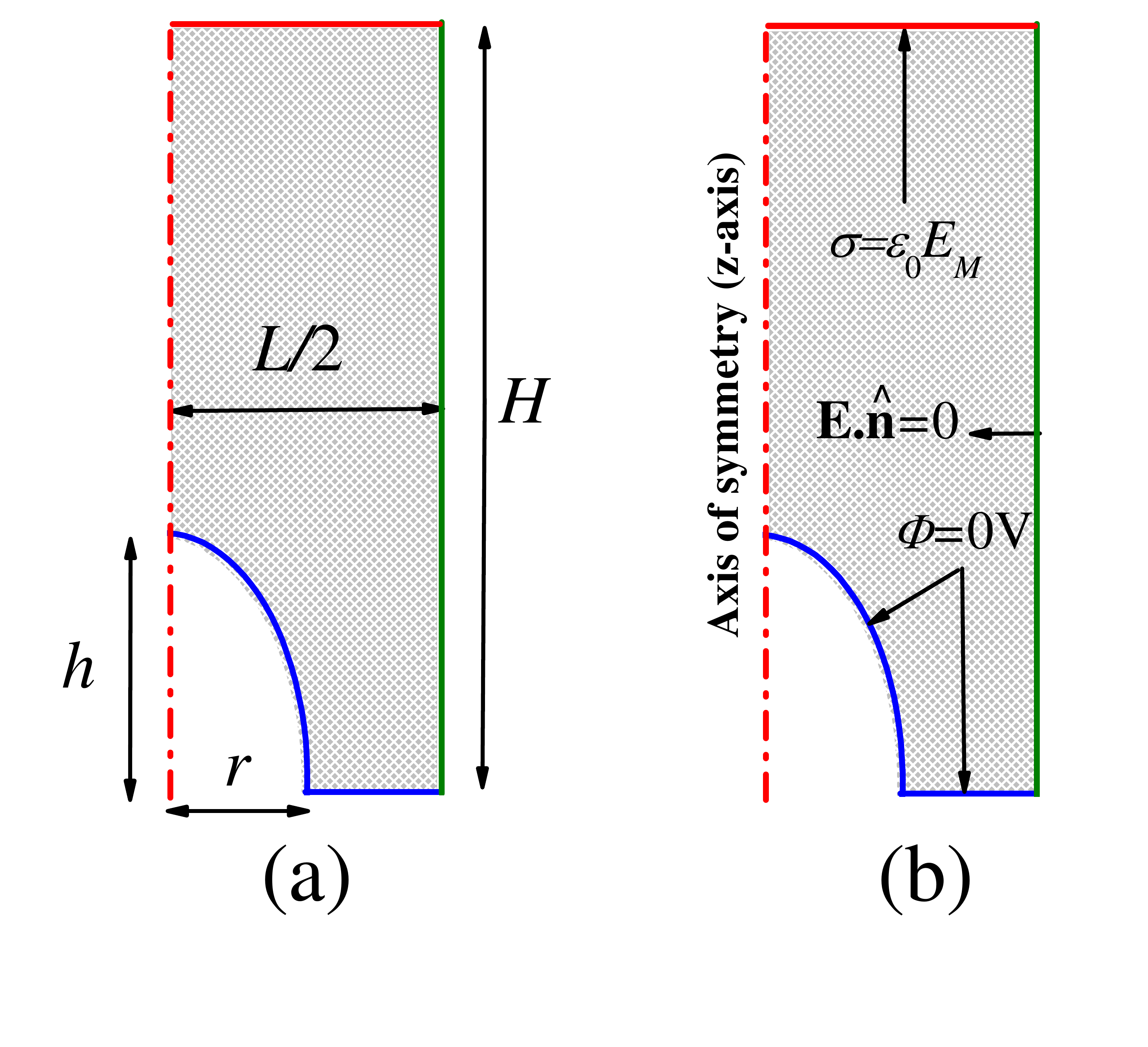}
\caption{Representation of simulation domain (gray color) and emitter with the corresponding dimensions indicated in (a) and the boundary conditions indicated in (b). The top boundary is a Neumann boundary condition that imposes a vertically aligned electric field as if the counter-electrode were at infinity.} \label{Hemi}
\end{figure}

The apex-FEF for HCP emitters have been extensively studied theoretically, although there are still several nuances to be investigated, especially for the variable $\delta^{n}$.
The numerical precision in $\gamma_{1}^{n}$ sensibly affects $\delta^{n}$, which is not well understood. Some authors using finite elements \cite{Edgcombe2001,Jo} had reported FEF with error of $\sim 3\%$ in the range of aspect ratios $4 \leqslant h/r \leqslant 3000$ ($r$ corresponds to the radius of the hemisphere over the cylindrical post). This error may be sufficiently small for many analyses. However, errors of this order can influence the true behavior in $\delta$ for large $c$, as we will discuss below. Furthermore, an analytical solution for HCP emitters is not known nor the existing formulae are particularly good to fit the whole range of aspect ratios including very low and very high values. The numerical works by Edgcombe and collaborators based in finite elements method \cite{Edgcombe2001,Edgcombe}, Kokkorakis and collaborators simulating a HCP as a cylindrical array of touching spheres \cite{Xanthakis} and Read and Bowring based on boundary elements method \cite{RBowring}, corroborates these comments. More recently, some works have analyzed numerically, by using finite elements, the problem of apex-FEF in single HCP emitters \cite{ZENG2009,Unicamp2016}. However, detailed information about the dimensions of the domain of simulation was not given, and the issue about how close the numerical results were to the analytical ones was not answered. Thus, we investigate the electrostatics of two identical hemi-ellipsoidal emitters, where the corresponding apex-FEF is known exactly for an isolated single emitter \cite{Edgcombe}. Let $r$ denote the semi-minor axis of the generating ellipse and $h$ its semi-major axis, i.e., the ellipses' radius and height are $r$ and $h$, respectively (see Fig.\ref{Hemi}), and define:

\begin{equation}
\nu \equiv h/r \hspace{0.3cm} ; \hspace{0.3cm} \xi \equiv \left(\nu^2 - 1\right)^{1/2}.
\label{hemiellipsoid}
\end{equation}
The apex FEF of a single hemi-ellipsoid emitter is then given by \cite{Edgcombe}:

\begin{equation}
\gamma_{1} = \frac{\xi^{3}}{\left[\left\{\nu\ln{(\nu+\xi)}\right\} - \xi \right]}.
\label{hemiellipsoid2}
\end{equation}
From Eq.(\ref{hemiellipsoid}), it is clear that for $\nu\rightarrow1$, $\xi\rightarrow0$, the hemi-ellipsoid becomes a hemisphere, and Eq.(\ref{hemiellipsoid2}) evaluates to $3$, as required \cite{Edgcombe}.

When doing computer simulation based on finite elements or finite differences, there are three main sources of errors: (i) the solution converges with low precision goal; (ii) mesh is too coarse or (iii) the simulation domain does not represents the intended system accurately. In this work, we focus on the latter item in the functional dependence of $\delta^{n}(c)$ as $c \rightarrow \infty$. Particularly, we investigate how the size of the simulation domain affects the error, regardless of the tolerance in numerical precision or the fineness of the mesh. Simulations that involve the solution of Laplace's equation using finite elements or finite differences require a minimum volume surrounding the region of interest. If domains are too small, the boundaries have an electrostatic effect on the emitter causing an undervaluation of the FEF. To avoid this issue, it may be possible to overestimate the size of the domain, however, this procedure may also be time and memory consuming, mainly in full three-dimensional systems. We define an error function $\varepsilon$ that is the relative difference between numerical and analytical values [via Eq.(\ref{hemiellipsoid2})] of the FEF:

\begin{equation}
\varepsilon(\%) = \frac{|\gamma_{1}^{n}-\gamma_1|}{\gamma_{1}} \times 100.
\label{hemiellipsoid3}
\end{equation}
\begin{figure}[h!]
\includegraphics [width=6.6cm,height=5.0cm] {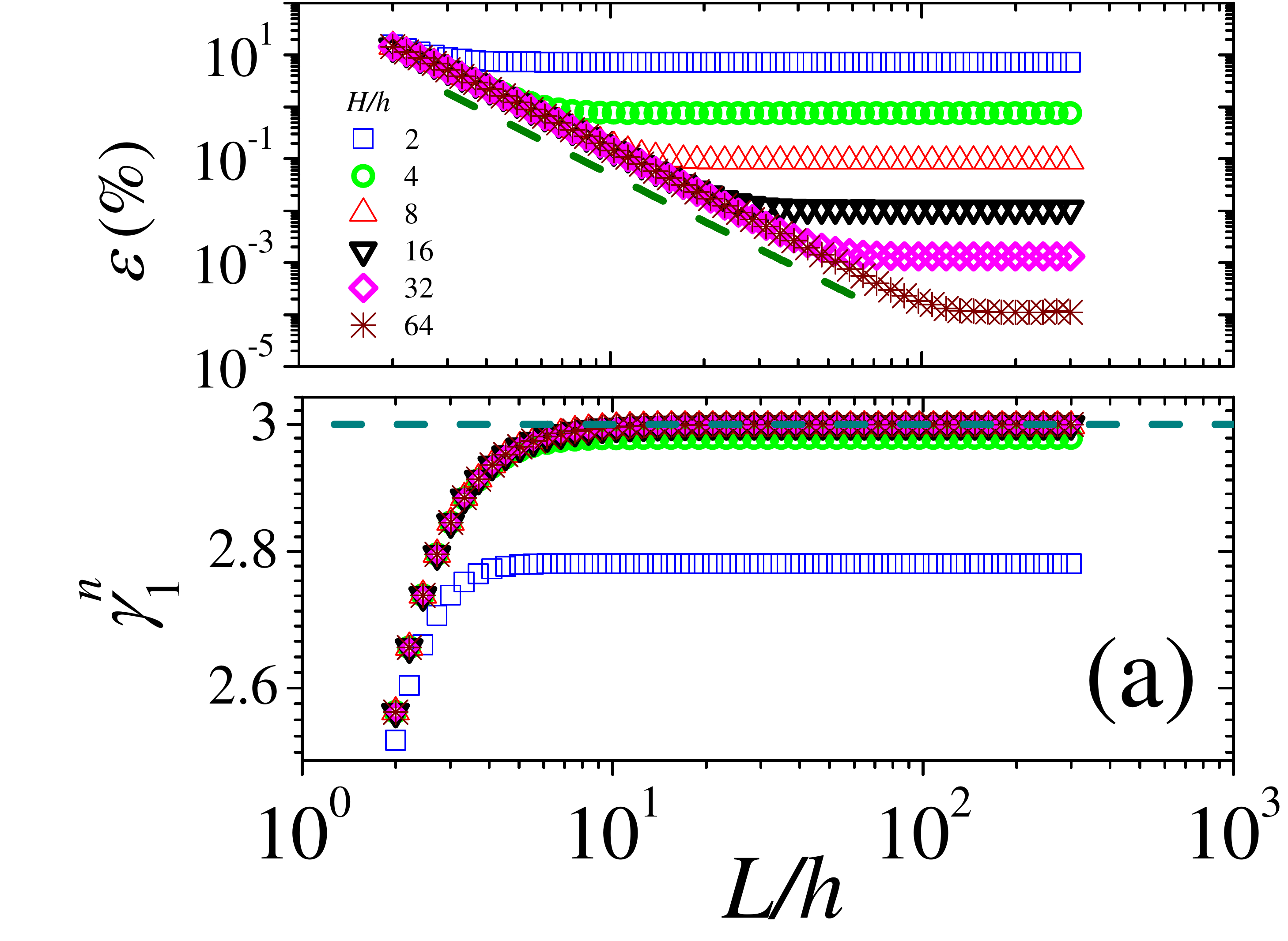}
\includegraphics [width=6.6cm,height=5.0cm] {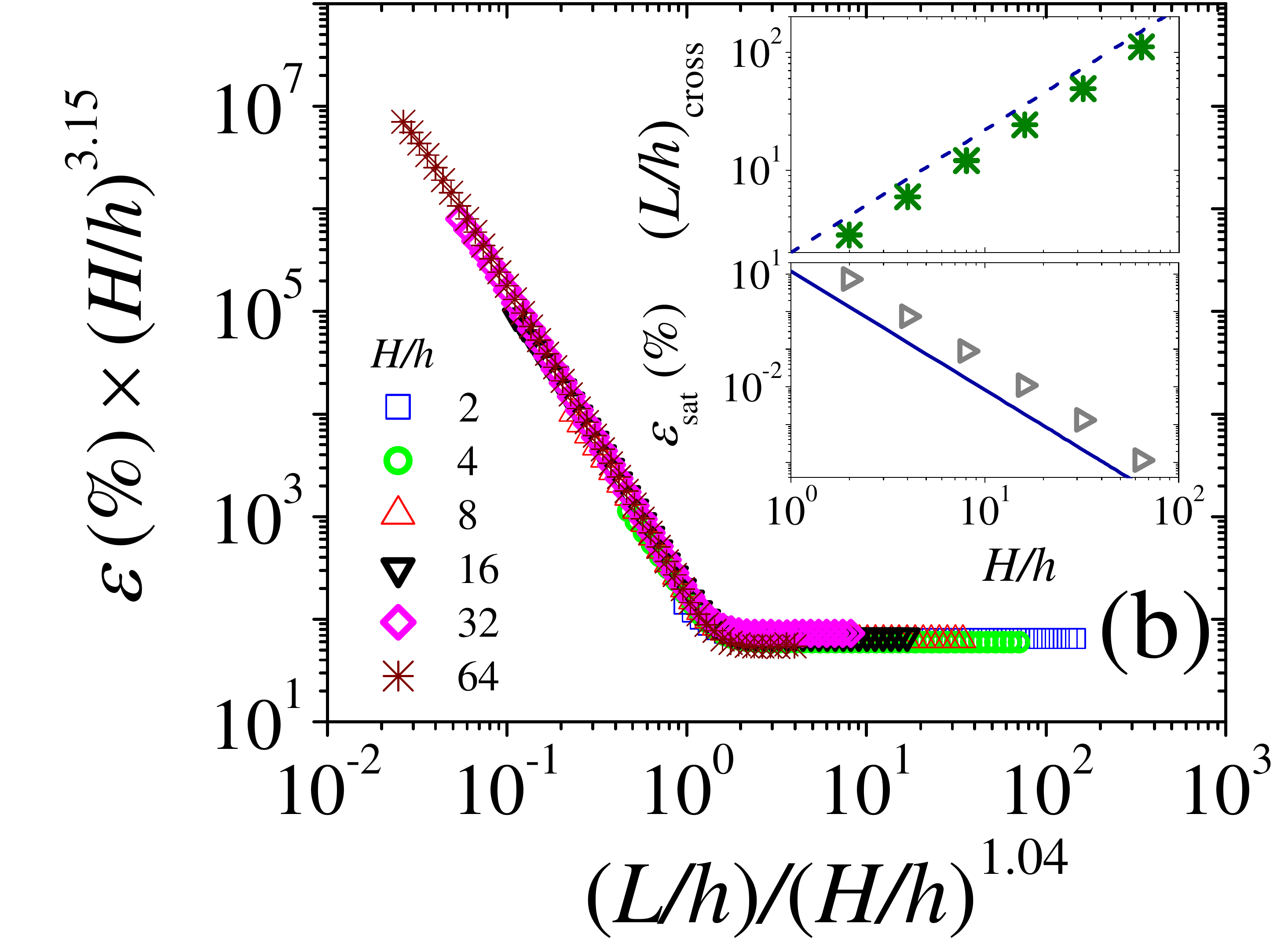}
\includegraphics [width=6.6cm,height=5.0cm] {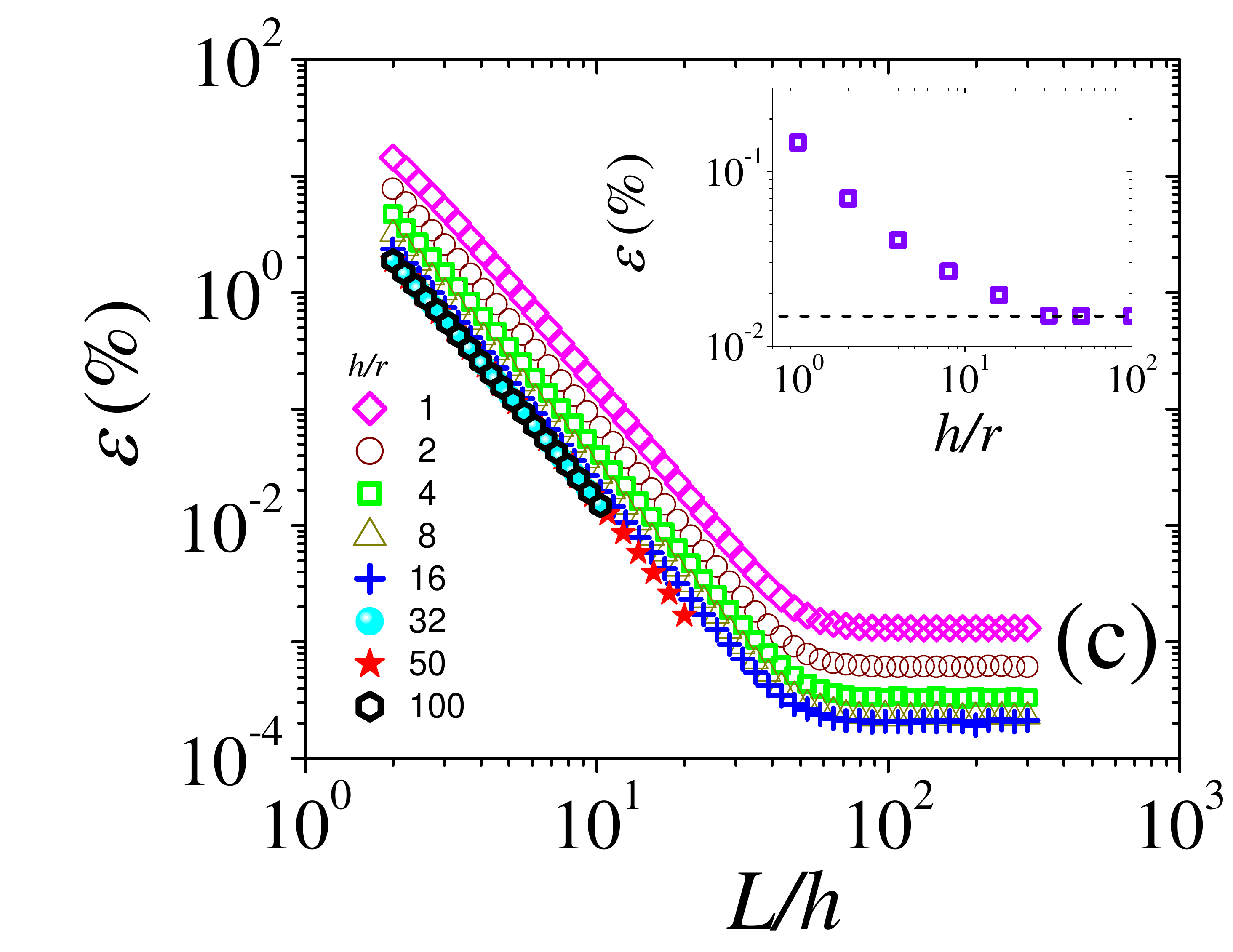}
\caption{(a) Error (top panel), $\varepsilon$ [see Eq.(\ref{hemiellipsoid3})], in the apex field enhancement factor of a hemisphere ($h/r=1$) due to the normalized size of the simulation domain $L/h$ for various values of the $H/h$. The slope of the dashed line is $-3$. The bottom panel shows the convergence of the numerical apex FEF toward the exact value (dashed line) for values of $H/h$ shown in the top panel. (b) Collapse of all curves shown in (a) by replacing the variables $\varepsilon\rightarrow\varepsilon\times(H/h)^{3.15}$ and $L/h\rightarrow \left(L/h\right) \times \left(H/h\right)^{1.04}$ (see text for details). The inset (top panel) shows the crossover ratio, $L/h_{cross}$, estimated from (a), as a function of $H/h$. The dashed line has slope ($1.04\pm0.01$). The bottom panel in the inset shows the saturation error, $\varepsilon_{sat}$, estimated from (a), in the plateau region, as a function of $H/h$. The full line has slope ($3.15\pm0.03$). (c) $\varepsilon$ as a function of $L/h$, for $H/h = 32$, and considering various aspect ratios. The inset shows the asymptotic decreasing of $\varepsilon$ as the aspect ratio increases for $L/h\approx 10.3$.} \label{Error}
\end{figure}

Figure \ref{Hemi}(a) represents the geometry of the physical system, and we are interested in knowing the sizes $L$ and $H$ to provide a desirable precision on $\gamma_{1}^{n}$. It is a two dimensional axisymmetric system with a central ellipsoidal emitter. We assume the emitter to be perfectly conductive (no electric field penetration); hence, the interior of the emitter is removed from the simulation domain. Figure \ref{Hemi}(b) shows the boundary conditions (BC). The emitter's surface and the bottom surface (substrate) are grounded (i.e., the electrical potential $\Phi$=0 V). The right hand side boundary is a symmetry line, i.e., this BC imposes that the electric field, $\textbf{E}$, is perpendicular to normal vector, $\textbf{\^{n}}$, from this boundary line (i.e., $\textbf{E}.\textbf{\^{n}}$=0) \cite{Fuzinato}. The top boundary is set as a surface charge density  $\sigma = \epsilon_{0}E_{M}$, where $\epsilon_0$ is the permittivity of vacuum and $E_{M}$ is the macroscopic electric field entering the domain, as appeared in Eq.(\ref{FEFa}). This BC assumes that the counter-electrode is at infinity, provided that $H$ is sufficiently large. The width of the simulation domain is, as usually defined in literature, $L/2$.

To calculate the FEF $\gamma_{1}^{n}$, we have used the commercial software COMSOL(version 5.3), based on the finite elements method. The details of the simulations are shown in the supplementary information. We want to stress that the analysis in this work is not limited to any particular method or computer code. Any method will require a minimum domain size to yield a desired precision. In our analysis, we took care to have enough numerical precision and sufficient number of mesh elements not to compromise the evaluation of $\varepsilon$. The $\varepsilon$ shall be solely due to the finite size of the simulation domain. Figure \ref{Error}(a) (top panel) shows $\varepsilon$ as a function of $L/h$, for several ratios $H/h$ and aspect ratio $h/r=1$ (i.e., a hemisphere emitter as discussed previously). It is interesting to observe (for example, for $H/h=64$) a very clear power law behavior showing that $\varepsilon$ scales as $\varepsilon \sim \left(L/h\right)^{-3}$ before saturation. This is a consequence of the electrostatic interactions between the emitter and its images generated by the boundaries. The exponent $-3$ that appears in Fig.\ref{Error}(a) indicates an universal trend in $\delta(c)$ for $c\rightarrow\infty$, which appears in other systems than Forbes' work about the FSEEP. The universality of the aforementioned trend will be detailed in section \ref{SecIII}.
The bottom panel illustrates the convergence of $\gamma_{1}^{n}$ towards $\gamma_{1}$ as the ratio $L/h$ increases, for same values $H/h$ shown in the top panel.

By exploring the scaling properties of electrostatics, it is possible to collapse all curves shown in Fig.\ref{Error}(a) (top panel) in a single curve, as shown in Fig.\ref{Error}(b). A good collapse observed reflects that the effects of $L$ and $H$ on $\gamma_{1}^{n}$ are scale invariant. The collapse of the data in a single curve was obtained in three steps. First, we estimate the value of the saturation error, $\varepsilon_{sat}$, by the plateau region from data presented in Fig.\ref{Error}(a) (top panel). A clear power law behavior is observed, where $\varepsilon_{sat} \sim \left(H/h\right)^{-3.15\pm0.03}$. Subsequently, a crossover at $\left(L/h\right)_{cross}$,  for each $H/h$, was calculated by intersection between $\varepsilon_{sat}$ and the power law decay. A nearly linear dependence $\left(L/h\right)_{cross} \sim \left(H/h\right)^{1.04\pm0.01}$ has been found. From these two steps we collapse the data by replacing the variables $\varepsilon\rightarrow\varepsilon\times(H/h)^{3.15}$  and $L/h\rightarrow \left(L/h\right) \times \left(H/h\right)^{1.04}$. Thus, our results suggest that $\varepsilon/\varepsilon_{sat}$ is a function of $(L/h)/\left(L/h\right)_{cross}$ only, i.e.,

\begin{equation}
\varepsilon(\%) \sim \varepsilon_{sat} \Im{\left\{\frac{(L/h)}{\left(L/h\right)_{cross}}\right\}},
\label{scaling}
\end{equation}
where $\Im$ is a scaling function. The existence of the saturation errors, $\varepsilon_{sat}>0$, suggests that, in addition to finite-size effects, the numerical solution converged (independent with the ratio $L/h$) for a given $H/h$. Thus, it is necessary to estimate the value of the ratio $\varepsilon(\%)/\varepsilon_{sat}$ at the plateau in Fig.\ref{Error}(b). With this procedure and considering the scaling given by Eq.(\ref{scaling}), one is able to determinate the error expected from numerical simulations for a given $H/h$. We have found (including error bars)

\begin{equation}
\varepsilon(\%) \approx \left(H/h\right)^{-3.15\pm0.03}\times(65\pm5).
\label{scaling2}
\end{equation}
For example, considering $H/h = 8$, Eq.(\ref{scaling2}) predicts $\varepsilon \approx 0.092 \%$ that is in excellent agreement with the corresponding error shown in Fig.\ref{Error}(a), that is $0.089 \%$.

Figure \ref{Error}(c) shows $\varepsilon$ as a function of $L/h$, for $H/h = 32$ and various aspect ratios. The results show that $\varepsilon$ tends to decrease, in an asymptotic way, when the $h/r$ increases. In the inset, we observe this trend, for $L/h\approx 10.3$. These results show that $\varepsilon$, for $h/r=1$, is the upper bound of the error so that, keeping the proportions $H/h$ and $L/h$ in the simulations for $h/r>1$, $\varepsilon$ is expected to be lower than that evaluated by using Eq.(\ref{scaling2}). Also, the error converges for $h/r\gtrsim32$ about one order of magnitude lower than the error for $h/r=1$.

\begin{figure}[h!]
\includegraphics [width=6.2cm,height=5.0cm] {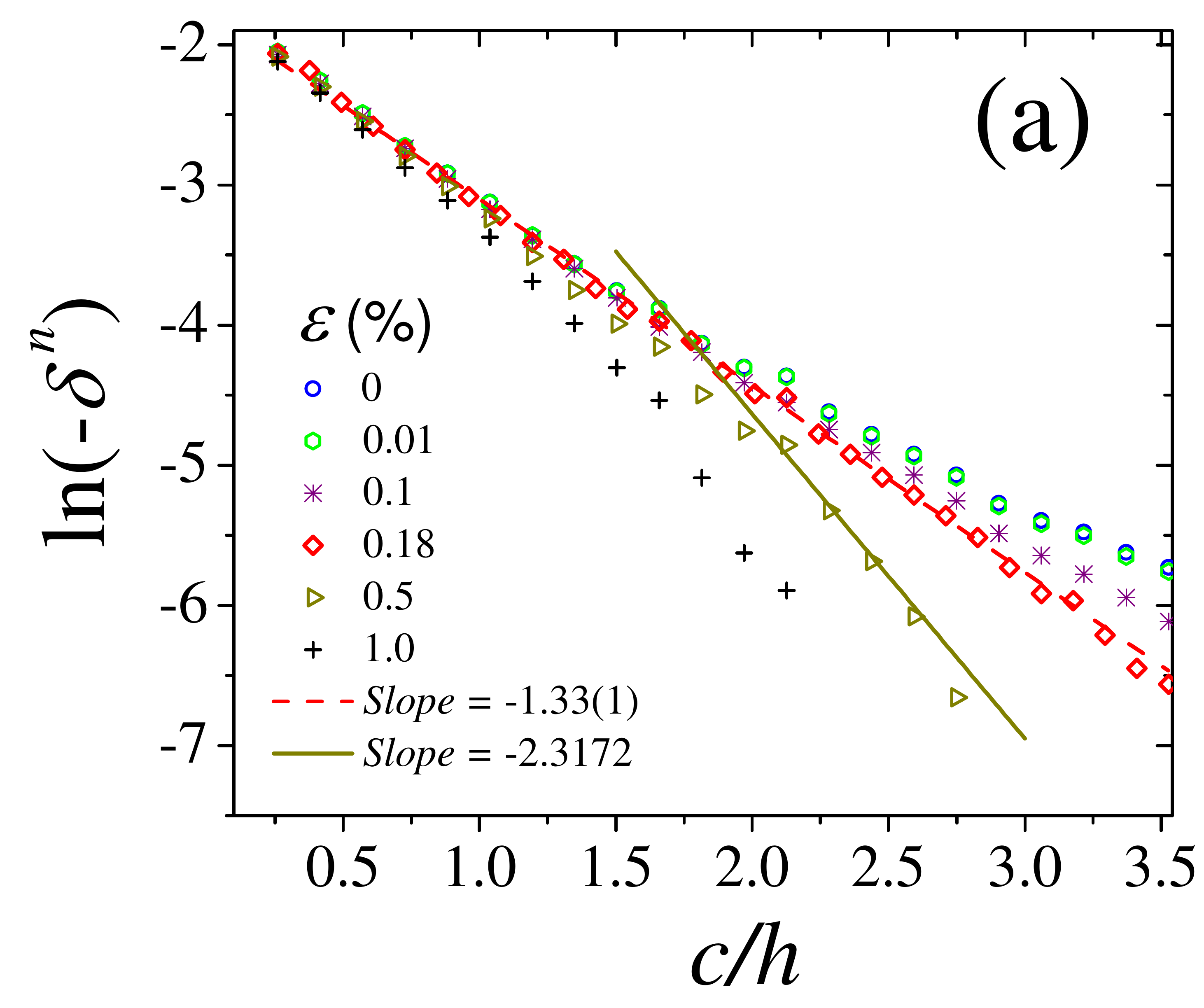}
\includegraphics [width=6.2cm,height=5.0cm] {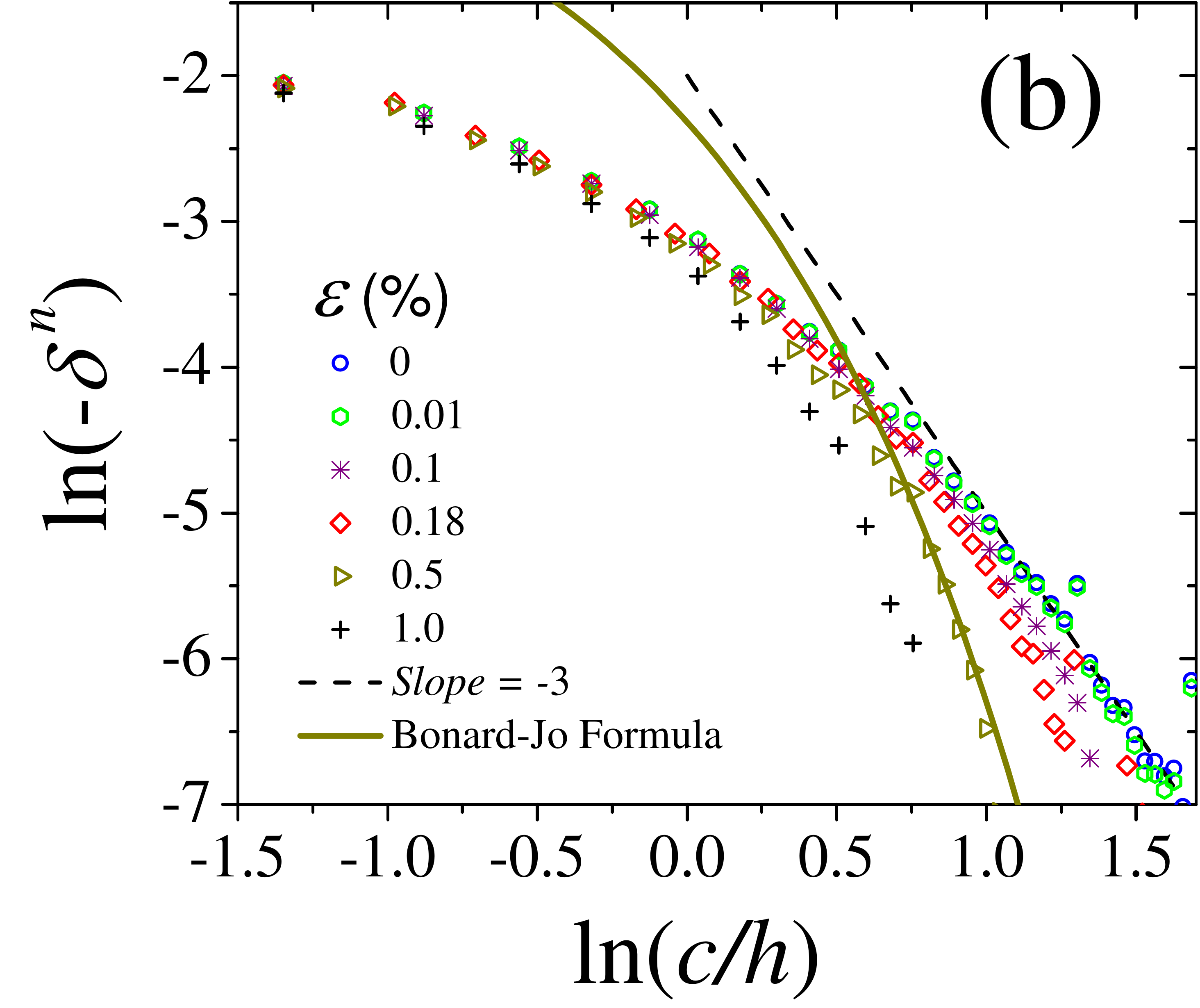}
\caption{Decay in $-\delta^{n}$ as a function of the normalized separation between the ellipsoids, $c/h$, for $h/r=8$. (a) Plots $\ln$-lin showing an apparent exponential decay for errors in  $\gamma_{1}^{n}$ of $0.18\%$. Slopes $(-1.33\pm0.01)$ (by least squares method) and $-2.3172$ (for comparison with Bonard-Jo results \cite{Bonard2001,Jo}) are also shown;  (b) Plots $\ln$-$\ln$ evidencing that only errors in $\gamma_{1}^{n}$ smaller than $0.01\%$ reveals the power law with decay $-\delta^{n} \sim (c/h)^{-3}$. Plot of Bornard-Jo formula [see Eq.(\ref{delta21Bonard})] and the slope $-3$ are also shown. Plots corresponding to other errors up to $1\%$ are also shown.} \label{Error5}
\end{figure}

\section{Fractional change in the apex-FEF for a two-identical hemi-ellipsoidal emitters case}
\label{SecIIa}

Next, we evaluate the apex-FEF from a pair of identical ellipsoidal emitters, $\gamma_{2}^{n}$, to see how the error in $\gamma_{1}^{n}$ is related to the trend in the $\delta^{n}$. For the numerical work, we use a full three-dimensional calculation, with the pair of emitters centered in a rectangular box. In this full 3D system containing a pair of emitters, the width of the domain in the direction perpendicular to the line containing the pair is $L/2$, whereas along the line that contains the pair the width must be $(L+c)/2$, so that the distance between an emitter and the closest symmetry boundary remains $L/2$. Hence, the dimensions of an axisymmetric system that generates an error is $L/2 \times H$, whereas the 3D domain that is compatible with the error must be $L/2 \times  (L+c)/2 \times H$. In fact, the error in the 3D system is expected to be slightly lower, as shown in Ref. \cite{Fuzinato}.

To determine $\delta^{n}$, we compute  $\gamma_{2}^{n}$ with error $\varepsilon=0.01\%$, using the size of the domain compatible with this error.
With this error, we can consider the $\gamma_{2}^{n}$ as exact, while $\gamma_{1}^{n}$ varies as a consequence of the variation in $L$ and $H$. The number of mesh elements used in these analysis were $\gtrsim 3\times 10^{5}$ depending on $L$ and $H$ and the convergence tolerance was $< 10^{-5}$. These choices are sufficient to avoid additional errors in the apex-FEFs, other than the error due to the finite domain size. The results are presented for $h/r=8$, which are expected to be the same for higher aspect ratios, as already observed in the inset of the Fig.\ref{Error}(c). Figure \ref{Error5} shows $\delta^{n}$  with different $\varepsilon$ in the determination of $\gamma_{1}^{n}$. Here, $\varepsilon=0$ mean that $\gamma_{1}^{n}=\gamma_{1}$, that is computed by Eq.(\ref{hemiellipsoid2}). In Fig.\ref{Error5}(a) $-\delta^{n}$  is plotted as a function of the ratio $c/h$, in mono-log plot, to evidence exponential behaviors as straight lines. The results indicate that the exact solution ($\varepsilon=0$) is clearly not an exponential. However, if $\gamma_{1}^{n}$ is computed with error of only $0.18\%$, $-\delta^{n}$ apparently decays exponentially with slope $-1.33\pm0.01$. The $\ln$-$\ln$ plot in Fig.\ref{Error5}(b) is best to review a power law decay trend for larger $c/h$. Indeed, the curve for $\varepsilon=0$ has the slope that trends to $-3$ as predicted by Forbes, using the FSEPP model \cite{RFJAP2016}, and by us with two HCP emitter's model \cite{ForbesAssis2017}. These results suggest that the physics behind charge-blunting effect manifests as power law behavior $-\delta \sim c^{-3}$ in the limit for $c\rightarrow \infty$.

Exponential decays in $-\delta$, have been hypothesized in the experimental works of Gr\"{o}ning \textit{et al.} \cite{Groning2000}, that used carbon nanotube (CNT) thin films deposited by a plasma
enhanced chemical vapor deposition process, and Cole \textit{et al.} \cite{Cole} with Large Area Field Emitters (LAFEs) formed by carbon nanofibres. Similar exponential decay rates (slopes close to $-1.40$ and $-1.25$, respectively) were obtained within the experimental error. We observe exponential behavior with similar slope for $\varepsilon=0.18\%$, possibly justifying Gr\"{o}ning's and Cole's results by the error in the $\gamma_1$ estimation (close to $500$ in Fig.9(a) of the Ref.\cite{Groning2000} and $300$ in Fig.2(c) of the Ref.\cite{Cole}). Similarly, Bonard \textit{et al.} and Jo \textit{et al.} \cite{Bonard2001,Jo} have found exponential decays in $-\delta^{n}$ for patterned
CNT films, consistent with errors in $\gamma_{1}^{n}$ of the order of $\varepsilon = 0.5\%$, as shown in Fig.\ref{Error5}(a), where slope similar to that observed by Eq.(\ref{delta21Bonard}) (i.e., $-2.3172$) has been found. As a consequence, in the high $c/h$ limit, slopes larger than $-3$ in $\ln$-$\ln$ plot are observed, as shown in Fig.\ref{Error5}(b). Errors smaller than $1\%$ can be considered good for many situations, however, to determine the asymptotic trend in $-\delta^{n}$, the evaluation of the apex-FEF of the emitters must be carried out with high precision ($\varepsilon \lesssim 0.01\%$, say) where the numerical and analytical results are indistinguishable, as shown in Fig.\ref{Error5}.

\section{Evidence of universality and signature of charge blunting effect in clusters or arrays}
\label{SecIII}

\begin{figure}[h!]
\includegraphics [width=6.6cm,height=5.0cm] {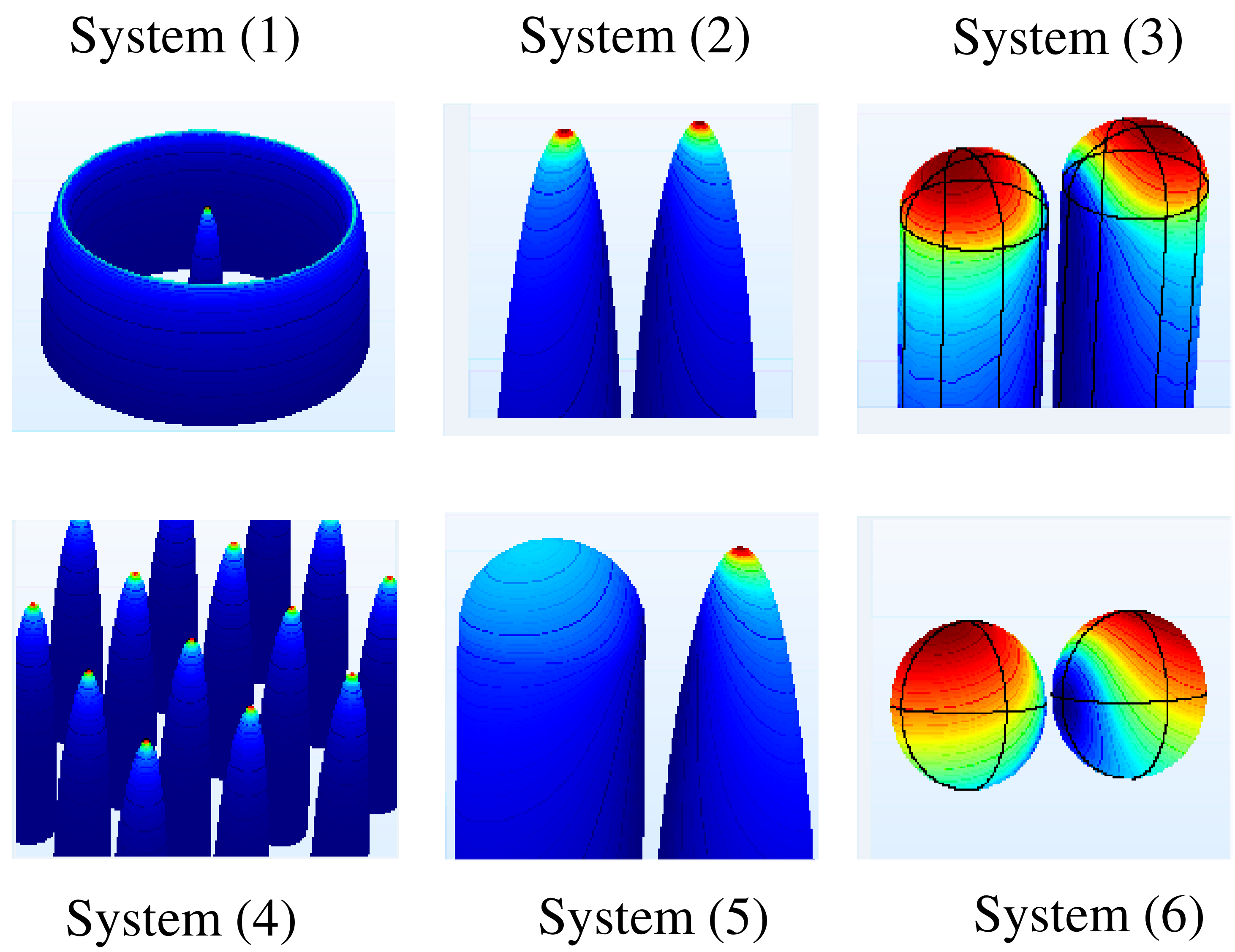}
\caption{Systems studied: (1) a ring with a hemi-ellipsoid emitter at the center; (2) two identical hemi-ellipsoids; (3) two identical HCP emitters; (4) infinite square array formed by identical hemi-ellipsoidal emitters; (5) two different emitters formed by a hemi-ellipsoid and a HCP; (6) two identical floating spheres. The color map represents the local FEF [red (blue) color indicates higher (lower) local FEF].} \label{Systems}
\end{figure}

As discussed in Ref.\cite{RFJAP2016}, the decreasing of the local charge at the apex of an emitter and the consequent decrease of the corresponding local FEF, involve charge transfer between the emitter and the conducting plate. This phenomenon is called charge blunting (CB) effect. This well known effect occurs due to approximation between emitter and its neighbors in a cluster or array.
In this section, we show strong evidence that the functional relation $-\delta^{n} \sim c^{-3}$, at sufficient large distances, is universal for emitters with any shape in small clusters or arrays. We consider here six physical systems, namely: (1) a ring with a hemi-ellipsoid emitter at the center; (2) two identical hemi-ellipsoids; (3) two identical HCP emitters; (4) an infinite square array formed by identical hemi-ellipsoidal emitters; (5) two different emitters formed by a hemi-ellipsoid and a HCP and (6) two identical floating spheres. All emitters and plates are assumed to be conductors. The system (1) is the 3D equivalent of the 2D axisymmetric system in Fig.\ref{Hemi} formed by a semi-ellipse and its image with respect to the right-hand boundary explained in Sec.\ref{SecII}. Thus, if the system is rotated around the symmetry axis of the semi-ellipse, the latter becomes a 3D hemi-ellipsoid and its image resembles a ring. If we do the transformation $\gamma_{1}^{n} \rightarrow \gamma_{2}^{n}$ and $\varepsilon(\%)/100 \rightarrow -\delta^{n}$ in Eq.(\ref{hemiellipsoid3}), we are able to perform the same analysis of a two-emitter system (a ring and a hemi-ellipsoid at its center) in the same way of that performed on Sec.\ref{SecIIa}, but for a central emitter. System (1) is advantageous in several aspects, since axysimmetric systems are easier to model, computation is fast and stable, the memory requirement is much less, the mesh can be made sufficiently fine and, therefore, numerical errors are smaller \cite{Fuzinato}. Furthermore, the values of $L$ and $H$ required to provide a given error in the axisymmetric system, shall provide a smaller error in a realistic full 3D model. That is, the 2D axisymmetric system always overestimate the error due to the finite size of the domain, granting the error to be within the desired tolerance. The systems (2) and (3) are full three dimensional system as that shown in Sec.\ref{SecIIa}.
Of course, a three dimensional system requires a comparatively large computing memory and are more time consuming. So, they are generally restricted to a limited range of the spacing between the emitters in a cluster or arrays. The results of the system (6) are presented considering the analytical formulas for fractional change in the apex-FEF by using Eqs.(17a) and (24) of Ref.\cite{RFJAP2016}. The systems (1)-(6) are illustrated in Fig.\ref{Systems} to provide a good insight.

\begin{figure}[h!]
\includegraphics [width=6.6cm,height=5.0cm] {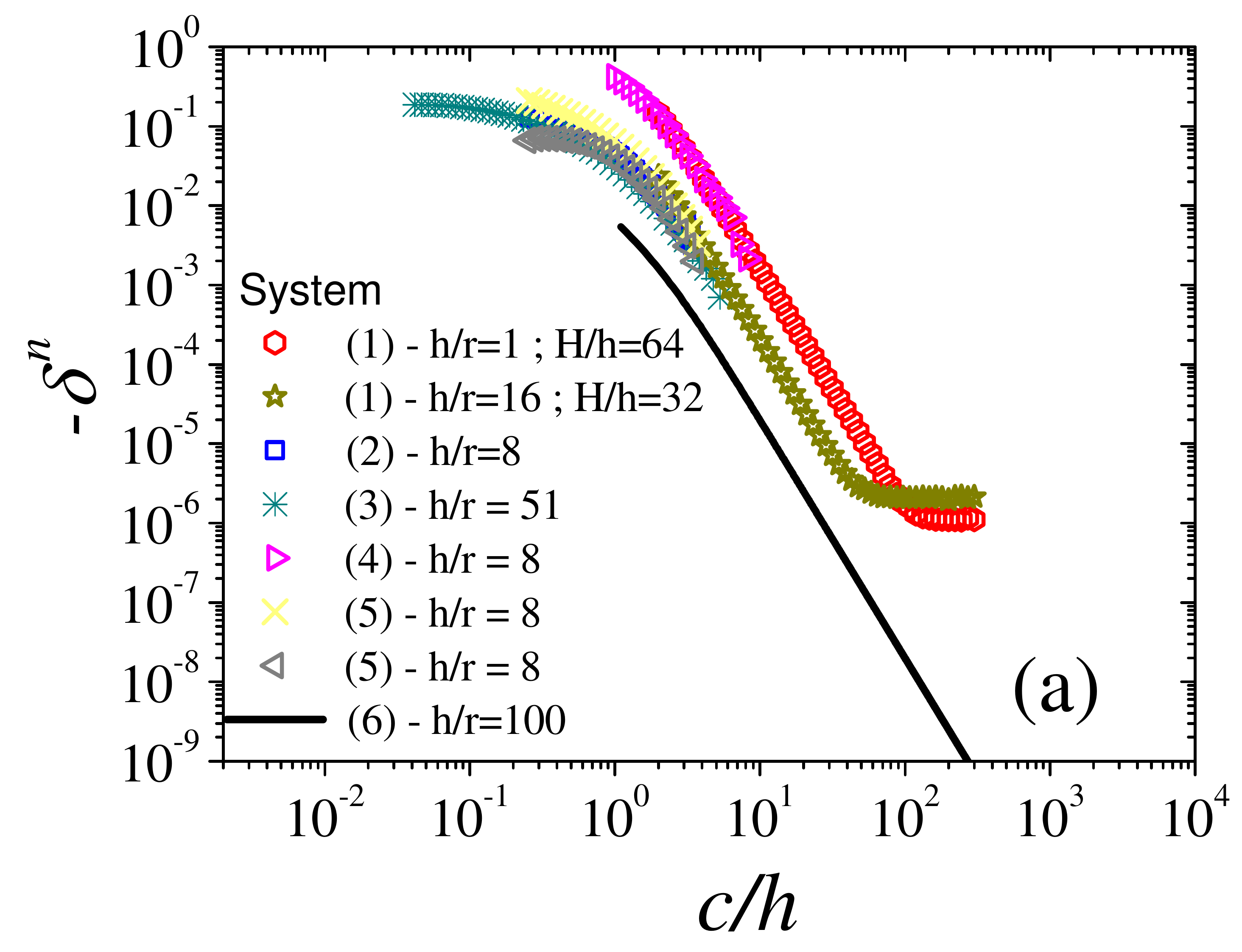}
\includegraphics [width=6.6cm,height=5.0cm] {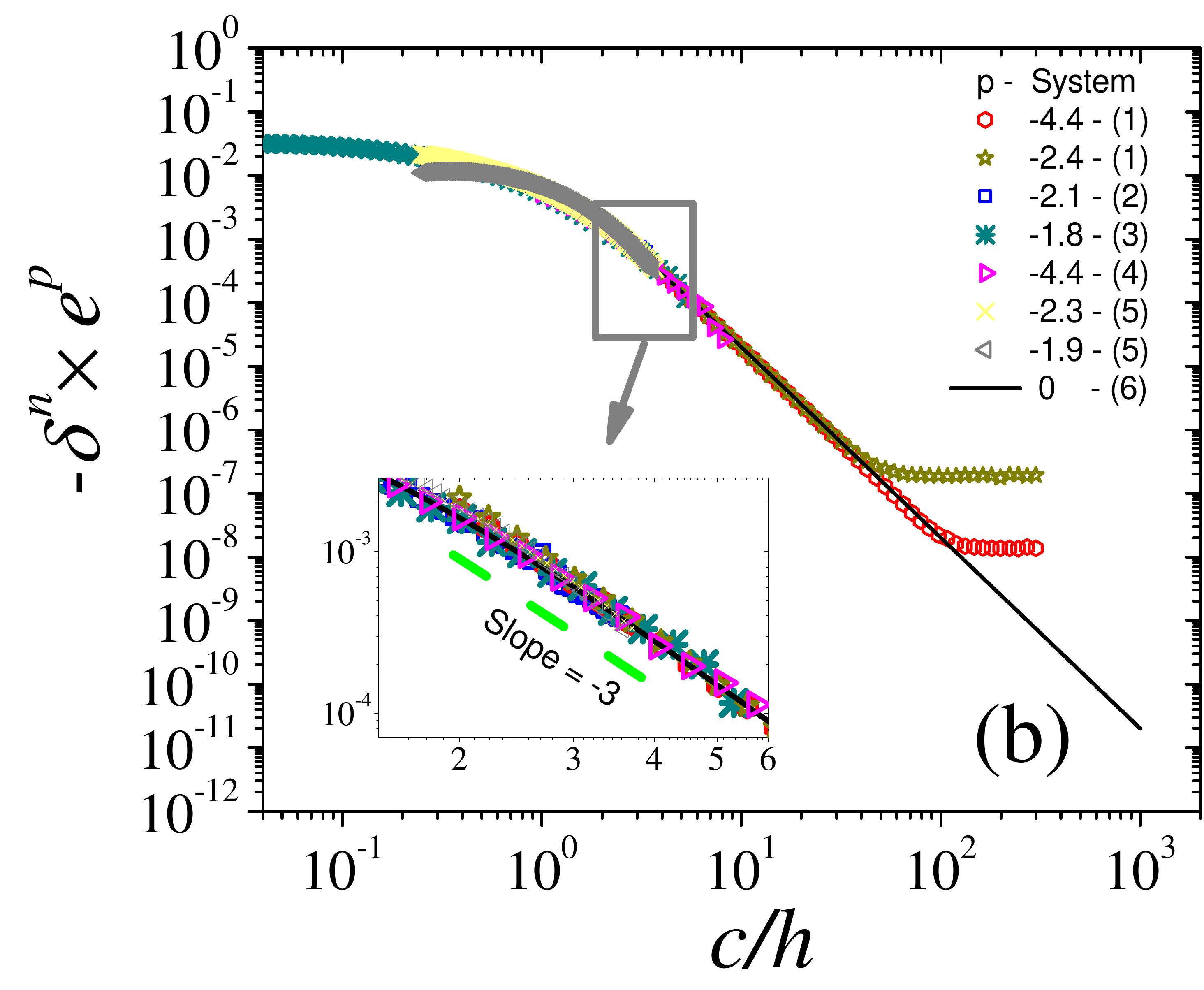}
\includegraphics [width=6.6cm,height=5.0cm] {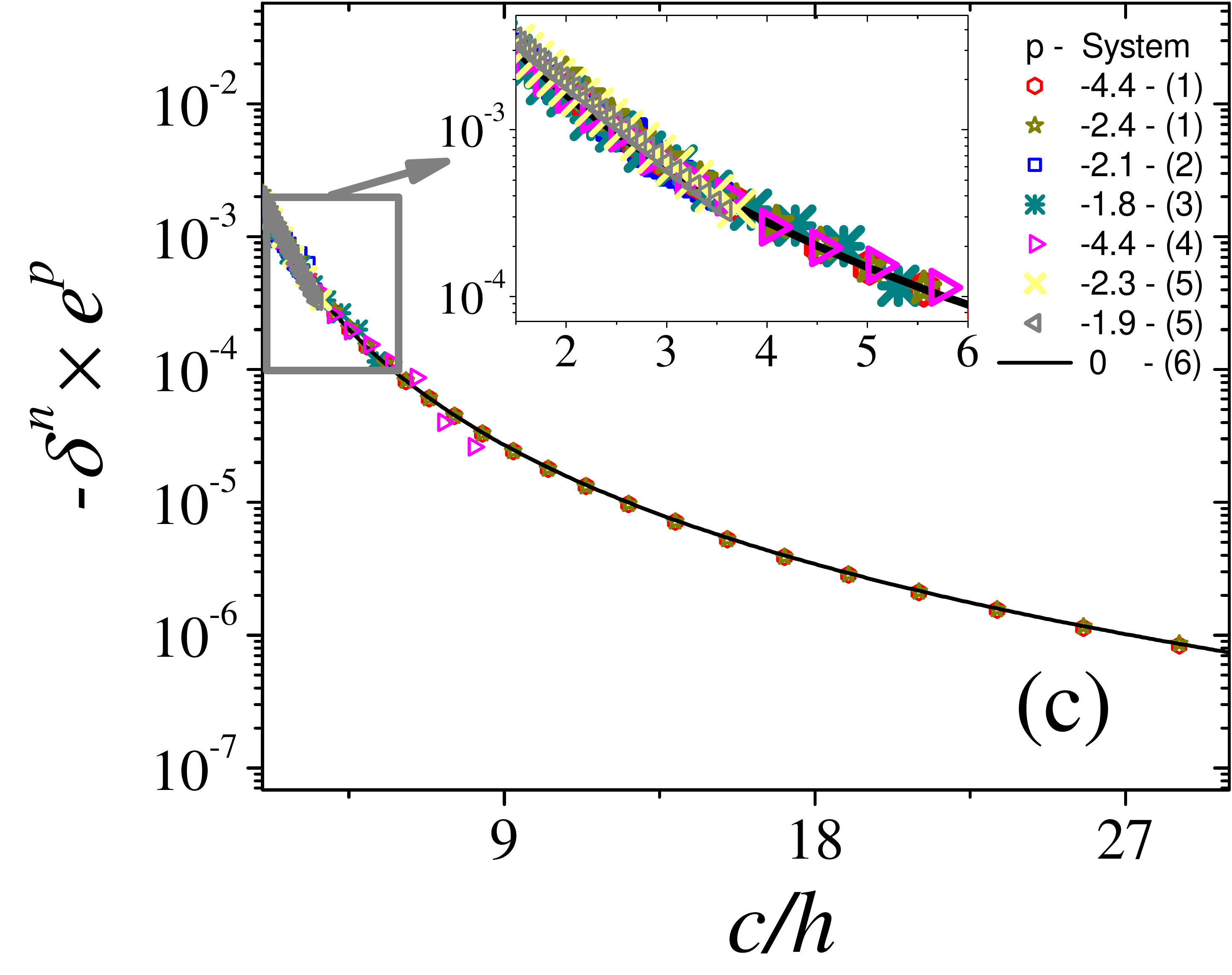}
\caption{(a) Numerical results of $-\delta^{n}$, as a function of ratio $c/h$, in $log-log$ scale, for the systems (1)-(6), shown in Fig.\ref{Systems}. (b) Results of the transformation $-\delta^{n} \rightarrow \exp{\left(p\right)} \times -\delta^{n}$ in the data shown in (a). The values of $p$ and the corresponding system are shown in the legend. The inset shows the results of the main panel magnified for a typical range $1.5 \leqslant c/l \leqslant 6$. The slope $-3$ is also highlighted. (c) Same data plotted in (b) in log-lin scale.} \label{Error22}
\end{figure}

Figure \ref{Error22}(a) shows the numerical results of the fractional change in the apex-FEF $-\delta^{n}$, as a function of ratio $c/h$ in $log-log$ scale, for the systems (1)-(6). These systems have been analyzed under a variety of different shapes, aspect ratios and $H/h$ ratios. Interestingly, the $-\delta$ for arrays are relatively larger, as compared with those from small clusters, due to the larger electrostatic effects, for a given $c/h$ value. It is worth to note that, for $c/h \rightarrow \infty$, the power law regime,

\begin{equation}
 -\delta^{n} = K \left(\frac{c}{h}\right)^{-3},
 \label{powerlaw}
\end{equation}
is the asymptotic behavior expected for all systems studied, either finite clusters or infinite arrays, either peripheral emitters or central ones. The corresponding values of $K$, depending on the system, account for the fact that $-\delta^{n}$ functions are vertically translated, when we observe the behavior for different systems. In Fig. \ref{Error22}(b), the results of the transformation $-\delta^{n} \rightarrow -\delta^{n} \times \exp{\left(p\right)}$, where $p$ is dependent on the system, are shown. This transformation corresponds to a vertical translation in the logarithmic scale. We considered the system (6) as a reference system, i.e., $p=0$, since the corresponding results showed in Fig.\ref{Error22}(a) were the analytical ones found in Ref.\cite{RFJAP2016}. For each system, the corresponding values of $p$ are shown in the legend of Fig. \ref{Error22}(b). For convenient values of $p$, all curves shown in Fig.\ref{Error22}(a) collapse suggesting that, in the limit of an infinite domain (i.e., if saturation effects in $-\delta^{n}$ curves are not verified), all curves tend to a single curve with slope $-3$ in a $log-log$ plot. In the inset of Fig. \ref{Error22}(b), the results of the main panel are magnified for a range $1.5 \leqslant c/h \leqslant 6$. We realize a clear tendency to a power law behavior for all systems studied. In contrast, for $c/h \lesssim 1.5$, the charge blunting is strong enough to alter the dependence of $-\delta^n$ with $c$, as already observed in the FSEPP model \cite{RFJAP2016}.

In Fig. \ref{Error22}(c), the same data plotted in Fig. \ref{Error22}(b) is shown in mono-log scale. If the dependence between $-\delta^{n}$ and $c$ were in fact exponential, the mono-log plot should be best to reveal the exponential dependence as straight lines. However, the curves are clearly not straight, showing that the fitting given by Eq.(\ref{delta21Bonard}) does not works for our systems. This aspect contrasts with results reported in literature by using finite elements or finite differences techniques. Finally, we stress that the generalization of Eq.(\ref{delta21Bonard}) proposed by Harris and collaborators by using LCM \cite{Harris2015AIP,Harris2016}, i.e., the Eq.(\ref{delta21Jensen}), is based in a two parameter fit equation, $a$ and $\kappa$, for a range similar to $1.5 \leqslant c/h \leqslant 6$, that configures a crossover region with a tendency to a asymptotic power-law behavior. In fact, Eq.(\ref{delta21Jensen}) showed to be a good fitting equation for this range with our data. However, Eq.(\ref{delta21Jensen}) does not predicts the asymptotic power-law dependence $-\delta^{n} \sim c^{m}$. Thus, our work shows that only one parameter equation, based in a power-law fitting, is sufficient with an universal exponent, $m=-3$, that is a signature of the beginning of the CB effect in emitters of a cluster or array.

Thus, we strongly suggest for accurate characterization of conducting emitters, in small clusters or arrays and at sufficiently large distances, to use the following one-parameter fit equation [easily derived from Eq.(\ref{powerlaw})] for apex-FEF influenced by CB effect:

\begin{equation}
\gamma_{CB} = \gamma_1 \left[1 - K\left(\frac{c}{h}\right)^{-3}\right].
 \label{powerlaw2}
\end{equation}
Given the generality of our results for small clusters or arrays, $\gamma_{CB}$ is used instead $\gamma_{2}$.

\begin{figure}[h!]
\includegraphics [width=6.6cm,height=5.0cm] {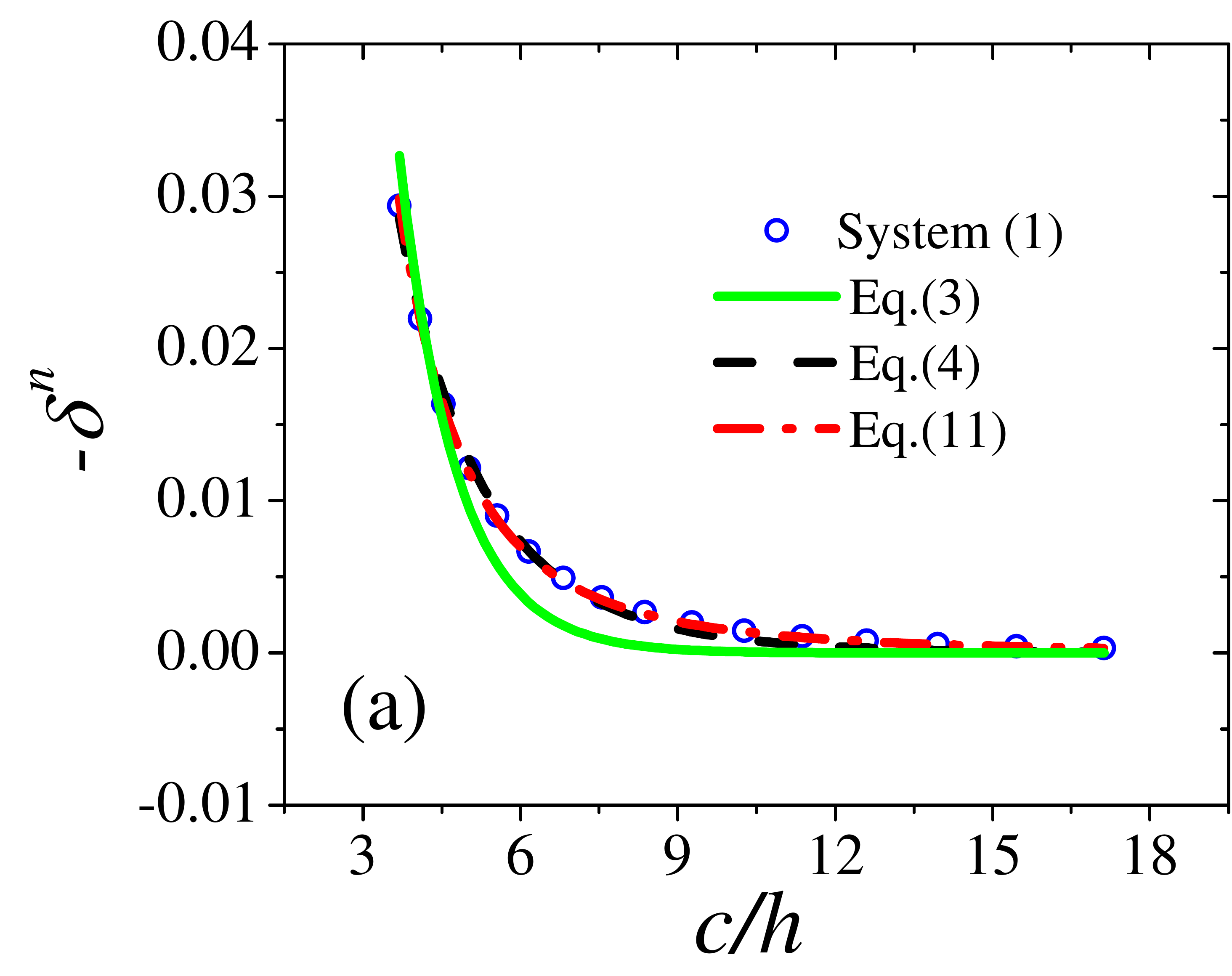}
\includegraphics [width=6.6cm,height=5.0cm] {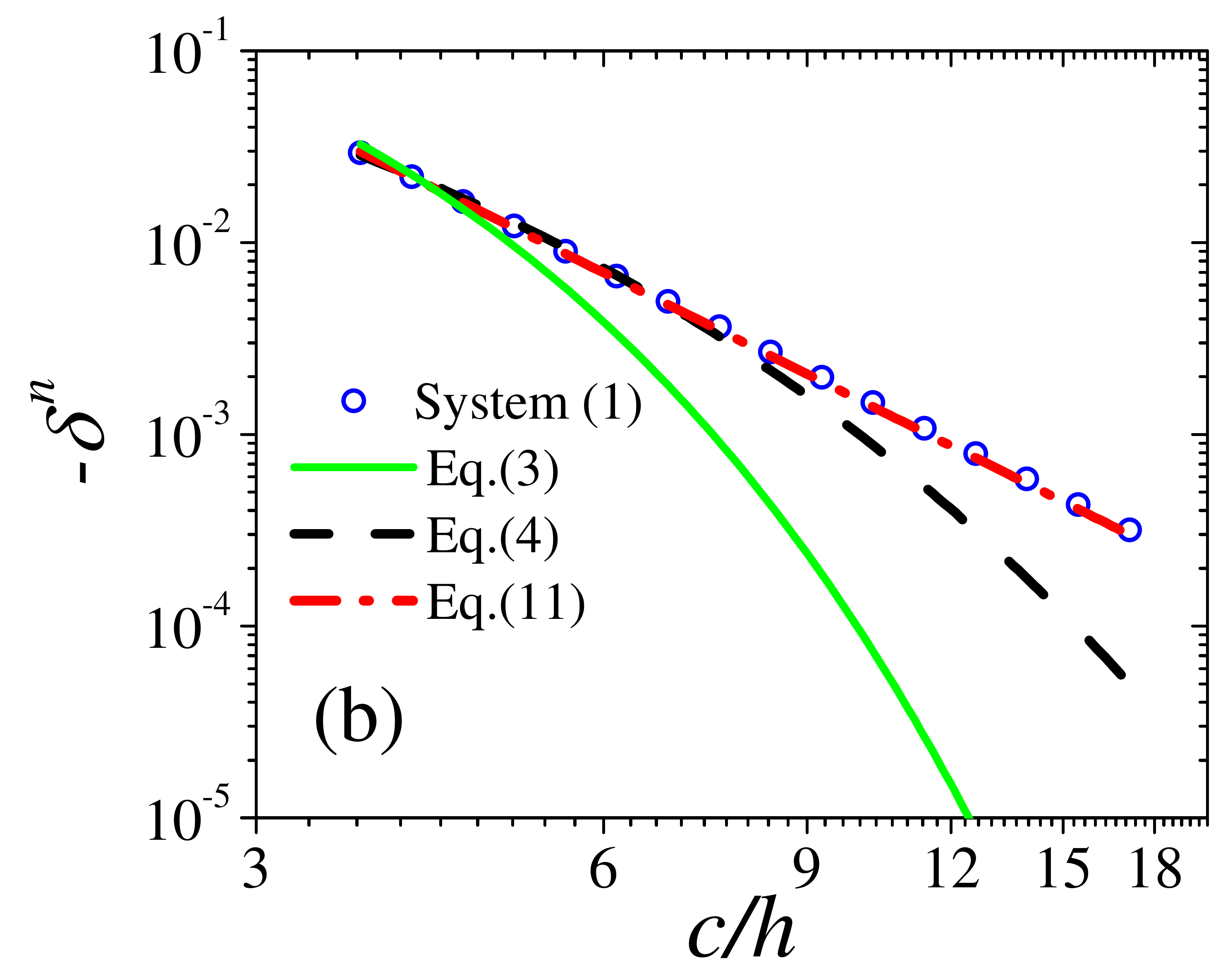}
\caption{Numerical results of $-\delta^{n}$ for the system (1), considering the range $3.5 \lesssim c/h \lesssim 17$ and the least square fittings by using Eq.(\ref{delta21Bonard}) with a free parameter [full green line], Eq.(\ref{delta21Jensen}) [dashed black line], and Eq.(\ref{powerlaw2}) [dashed dot red line]. The plots are shown in (a) lin-lin and (b) log-log scale.} \label{Systemsproof}
\end{figure}

Let us show numerically that, at sufficiently large distances, Eq.(\ref{powerlaw2}) is the best fit equation as compared with fittings given by Eqs.(\ref{delta21Bonard}) and (\ref{delta21Jensen}). For this, we have used the numerical results from system (1), where $-\delta^{n}$ was computed for a broad range of $c/h$. Figure \ref{Systemsproof} shows the numerical results for the system (1), considering the range $3.5 \lesssim c/h \lesssim 17$. Also, for fitting our numerical results, we have used three fitting equations, namely: (i) modified Eq.(\ref{delta21Bonard}) with a free parameter, $a$, that corresponds to Eq.(\ref{delta21Jensen}) with $\kappa=1$; (ii) Eq.(\ref{delta21Jensen}), and (iii) Eq.(\ref{powerlaw2}). Figure \ref{Systemsproof}(a) shows the numerical results and fittings in a lin-lin scale. For function (i), we have obtained by a least square fit $a=(-0.92\pm0.01)$ (different, therefore, of $-2.3172$) and adjust R-square$=0.93745$. For function (ii), we have obtained by a least square fit $a=(-1.48\pm0.03)$, $\kappa=(0.66\pm0.01)$  and adjust R-square$=0.99661$. For function (iii), we have obtained by a least square fit $K=(1.503\pm0.006)$ and adjust R-square$=0.99952$. Thus, the results obtained by using function (iii) provide more precise fitting, even with only one parameter. Figure \ref{Systemsproof}(b) shows the same results of Fig.\ref{Systemsproof}(a), in log-log plot, confirming the good agreement by comparison of the function (iii) and numerical data in at least two decades in the vertical scale. This feature was not observed for the other functions.

\section{Conclusions}

We have shown the origin of the literature discrepancies in the fractional reduction of the apex-FEF, $\delta$, considering field emitters in small clusters or arrays, by using precise numerical calculations. Our results show that the effects of the domain height, $H$, and lateral size, $L$, on $\varepsilon$ are scale invariant. This feature allows one to predict the error on the calculation of an apex-FEF of a single tip field emitter with specified ratios $L/h$ and $H/h$. This scaling relation allows one to work with a desired error for accurate study of the electrostatic effects on the fractional reduction of the apex-FEF, due to the proximity of the emitters in small clusters or arrays. Our results also suggest that the power law functional dependence, $-\delta \sim c^{-3}$, is universal for large distances between emitters, applicable to emitters with any shape in small clusters or infinite arrays and a signature of the charge-blunting effect in LAFEs. This contrasts with a long time established exponential decay, in which an apparent non-universal decay rate, as reported in the literature, seems to be an effect of error $0.2\%\lesssim\varepsilon\lesssim0.5\%$ in the calculation of $\gamma_{1}^{n}$. These results improve the scientific understanding of the field electron emission theory. The use of the scaling presented in Eqs.(\ref{scaling}) and (\ref{scaling2}) is suggested in the numerical simulations for accurate comparison, characterization and understanding of the physics behind real emitters in clusters or arrays. Finally, one fit parameter equation [i.e., Eq.(\ref{powerlaw2})] is enough to fit the apex-FEF of an emitter subject to charge blunting effect, at sufficient large distances. This equation is more simple and is supported by precise numerical simulations, as presented here, and analytical results \cite{RFJAP2016}.

\section{Acknowledgements}
TAdA acknowledges Royal Society under Newton Mobility Grant, Ref: NII60031. TAdA thanks Richard Forbes for illuminating discussions during a scientific visit to University of Surrey (U.K.), where this work was motivated.


\begin{thebibliography}{10}
\expandafter\ifx\csname url\endcsname\relax
  \def\url#1{{\tt #1}}\fi
\expandafter\ifx\csname urlprefix\endcsname\relax\def\urlprefix{URL }\fi
\providecommand{\eprint}[2][]{\url{#2}}

\bibitem{Muller1937}
M{\"u}ller E~W 1937 {\em Zeitschrift f{\"u}r Physik\/} {\bf 106} 541--550 ISSN
  0044-3328

\bibitem{Muller1951}
M{\"u}ller E~W 1951 {\em Zeitschrift f{\"u}r Physik\/} {\bf 131} 136--142 ISSN
  0044-3328

\bibitem{Muller1956}
M\"uller E~W and Bahadur K 1956 {\em Phys. Rev.\/} {\bf 102}(3) 624--631

\bibitem{Edgcombe}
Forbes R~G, Edgcombe C and Valdrè U 2003 {\em Ultramicroscopy\/} {\bf 95} 57 --
  65

\bibitem{Cole2015chapter}
Cole M~T, Mann M, Teo K~B and Milne W~I 2015 Chapter 5 - engineered carbon
  nanotube field emission devices {\em Emerging Nanotechnologies for
  Manufacturing (Second Edition)\/} Micro and Nano Technologies ed Ahmed W,
  and Jackson M~J (Boston: William Andrew Publishing) pp 125 -- 186 second
  edition ed ISBN 978-0-323-28990-0

\bibitem{PRE1}
Djurabekova F, Parviainen S, Pohjonen A and Nordlund K 2011 {\em Phys. Rev.
  E\/} {\bf 83}(2) 026704

\bibitem{PRLNanotube}
Liang S~D and Chen L 2008 {\em Phys. Rev. Lett.\/} {\bf 101}(2) 027602

\bibitem{PRL2}
Pascale-Hamri A, Perisanu S, Derouet A, Journet C, Vincent P, Ayari A and
  Purcell S~T 2014 {\em Phys. Rev. Lett.\/} {\bf 112}(12) 126805

\bibitem{PRB1}
Cabrera H, Zanin D~A, De~Pietro L~G, Michaels T, Thalmann P, Ramsperger U,
  Vindigni A, Pescia D, Kyritsakis A, Xanthakis J~P, Li F and Abanov A 2013
  {\em Phys. Rev. B\/} {\bf 87}(11) 115436

\bibitem{Jeffreys}
Jeffreys H 1925 {\em Proceedings of the London Mathematical Society\/} {\bf
  s2-23} 428--436

\bibitem{FowlerN}
Fowler R~H and Nordheim L 1928 {\em Proceedings of the Royal Society of London
  A: Mathematical, Physical and Engineering Sciences\/} {\bf 119} 173--181 ISSN
  0950-1207

\bibitem{Burgess}
Burgess R~E, Kroemer H and Houston J~M 1953 {\em Phys. Rev.\/} {\bf 90}(4)
  515--515

\bibitem{MG}
Murphy E~L and Good R~H 1956 {\em Phys. Rev.\/} {\bf 102}(6) 1464--1473

\bibitem{Forbes}
Forbes R~G and Deane J~H 2007 {\em Proceedings of the Royal Society of London
  A: Mathematical, Physical and Engineering Sciences\/} {\bf 463} 2907--2927

\bibitem{ForbesJPhysA}
Deane J~H~B and Forbes R~G 2008 {\em Journal of Physics A: Mathematical and
  Theoretical\/} {\bf 41} 395301

\bibitem{Forbes2013}
Forbes R~G 2013 {\em Proceedings of the Royal Society of London A:
  Mathematical, Physical and Engineering Sciences\/} {\bf 469} 20130271

\bibitem{Jensen2015}
Harris J~R, Jensen K~L and Shiffler D~A 2015 {\em Journal of Physics D: Applied
  Physics\/} {\bf 48} 385203

\bibitem{JensenAPL2015}
Harris J~R, Jensen K~L, Shiffler D~A and Petillo J~J 2015 {\em Applied Physics
  Letters\/} {\bf 106} 201603

\bibitem{Harris2015AIP}
Harris J~R, Jensen K~L and Shiffler D~A 2015 {\em AIP Advances\/} {\bf 5}
  087182

\bibitem{Harris2016}
Harris J~R, Jensen K~L, Tang W and Shiffler D~A 2016 {\em Journal of Vacuum
  Science \& Technology B, Nanotechnology and Microelectronics: Materials,
  Processing, Measurement, and Phenomena\/} {\bf 34} 041215

\bibitem{Jensen2016AIPA}
Tang W~W, Shiffler D~A, Harris J~R, Jensen K~L, Golby K, LaCour M and Knowles T
  2016 {\em AIP Advances\/} {\bf 6} 095007

\bibitem{RFJAP2016}
Forbes R~G 2016 {\em Journal of Applied Physics\/} {\bf 120} 054302

\bibitem{deAssis1}
de~Assis T~A and Dall'Agnol F~F 2016 {\em Nanotechnology\/} {\bf 27} 44LT01

\bibitem{deAssisJAP}
de~Assis T~A and Dall'Agnol F~F 2017 {\em Journal of Applied Physics\/} {\bf
  121} 014503

\bibitem{ForbesAssis2017}
Dall'Agnol F~F, de~Assis T~A and Forbes R~G 2017 Electrostatic effect on the
  characteristic field enhancement factors when two identical emitters are in
  close proximity {\em 30th International Vacuum Nanoelectronics Conference
  (IVNC)\/} pp 230--231

\bibitem{FT2017JPCM}
Dall'Agnol F~F and de~Assis T~A 2017 {\em Journal of Physics: Condensed
  Matter\/} {\bf 29} 40LT01

\bibitem{Jensen2017}
Harris J~R, Jensen K~L, Petillo J~J, Maestas S, Tang W and Shiffler D~A 2017
  {\em Journal of Applied Physics\/} {\bf 121} 203303

\bibitem{Bonard2001}
Bonard J~M, Weiss N, Kind H, Stöckli T, Forró L, Kern K and Châtelain A 2001
  {\em Advanced Materials\/} {\bf 13} 184--188 ISSN 1521-4095

\bibitem{Jo}
Jo S~H, Tu Y, Huang Z~P, Carnahan D~L, Wang D~Z and Ren Z~F 2003 {\em Applied
  Physics Letters\/} {\bf 82} 3520--3522

\bibitem{Refnew2}
Harris J~R, Jensen K~L and Shiffler D~A 2015 {\em Journal of Physics D: Applied
  Physics\/} {\bf 48} 385203

\bibitem{LCMnew}
Jensen K~L 2010 {\em Journal of Applied Physics\/} {\bf 107} 014905

\bibitem{SRZhang}
Zhang Z, Meng G, Wu Q, Hu Z, Chen J, Xu Q and Zhou F 2014 {\em Scientific
  Reports\/} {\bf 4} 4676

\bibitem{ZhuRef}
Zhu Y~W, Yu T, Cheong F~C, Xu X~J, Lim C~T, Tan V~B~C, Thong J~T~L and Sow C~H
  2005 {\em Nanotechnology\/} {\bf 16} 88

\bibitem{Maiti}
Maiti U~N, Maiti S, Das N~S and Chattopadhyay K~K 2011 {\em Nanoscale\/} {\bf
  3}(10) 4135--4141

\bibitem{Sheini}
Jamali-Sheini F, Patil K, Joag D~S and More M~A 2011 {\em Applied Surface
  Science\/} {\bf 257} 8366 -- 8372 ISSN 0169-4332

\bibitem{WangAPL}
Wang L, Gao F, Chen S, Li C and Yang W 2015 {\em Applied Physics Letters\/}
  {\bf 107} 122108

\bibitem{Groning1}
Nilsson L, Groening O, Emmenegger C, Kuettel O, Schaller E, Schlapbach L, Kind
  H, Bonard J~M and Kern K 2000 {\em Applied Physics Letters\/} {\bf 76}
  2071--2073

\bibitem{Groning2000}
Gröning O, Küttel O~M, Emmenegger C, Gröning P and Schlapbach L 2000 {\em
  Journal of Vacuum Science \& Technology B: Microelectronics and Nanometer
  Structures Processing, Measurement, and Phenomena\/} {\bf 18} 665--678

\bibitem{DAHL20003}
Dahl D~A 2000 {\em International Journal of Mass Spectrometry\/} {\bf 200} 3 --
  25 ISSN 1387-3806 volume 200: The state of the field as we move into a new
  millenium

\bibitem{Edgcombe2001}
Edgcombe C~J and Valdr\`{e} U 2001 {\em Journal of Microscopy\/} {\bf 203}
  188--194 ISSN 1365-2818

\bibitem{Xanthakis}
Kokkorakis G~C, Modinos A and Xanthakis J~P 2002 {\em Journal of Applied
  Physics\/} {\bf 91} 4580--4584

\bibitem{RBowring}
Read F and Bowring N 2004 {\em Nuclear Instruments and Methods in Physics
  Research Section A: Accelerators, Spectrometers, Detectors and Associated
  Equipment\/} {\bf 519} 305 -- 314 ISSN 0168-9002 proceedings of the Sixth
  International Conference on Charged Particle Optics

\bibitem{ZENG2009}
Zeng W, Fang G, Liu N, Yuan L, Yang X, Guo S, Wang D, Liu Z and Zhao X 2009
  {\em Diamond and Related Materials\/} {\bf 18} 1381 -- 1386 ISSN 0925-9635

\bibitem{Unicamp2016}
Roveri D, Sant'Anna G, Bertan H, Mologni J, Alves M and Braga E 2016 {\em
  Ultramicroscopy\/} {\bf 160} 247 -- 251

\bibitem{Fuzinato}
Dall'Agnol F~F and den Engelsen D 2013 {\em Nanoscience and Nanotechnology
  Letters\/} {\bf 5}

\bibitem{Cole}
Cole M, Teo K~B~K, Groening O, Gangloff L, Legagneux P and Milne W~I 2014 {\em
  Sci. Rep.\/} {\bf 4} 4840

\end{thebibliography}

\providecommand{\newblock}{}

\end{document}